%% file: paper.tex
\documentclass[sigplan,10pt,screen,authorversion,nonacm]{acmart}

\renewcommand\footnotetextcopyrightpermission[1]{}
\setcopyright{none}
\title{Automated Tensor Scheduling for Hybrid CPU-GPU LLM Inference on Consumer Devices}

\author{Yangyijian Liu}
\authornote{These authors contributed equally to this work.}
\affiliation{%
    \institution{School of Computer Science, Nanjing University}
    \city{Nanjing}
    \country{China}
}
\email{yyj.liu@smail.nju.edu.cn}

\author{Hongyi Ye}
\authornotemark[1]
\affiliation{%
    \institution{School of Computer Science, Nanjing University}
    \city{Nanjing}
    \country{China}
}
\email{231220150@smail.nju.edu.cn}

\author{Mingyang Li}
\affiliation{%
    \institution{School of Computer Science, Nanjing University}
    \city{Nanjing}
    \country{China}
}
\email{limingyang@smail.nju.edu.cn}

\author{Wu-Jun Li}
\affiliation{%
    \institution{School of Computer Science, Nanjing University}
    \city{Nanjing}
    \country{China}
}
\email{liwujun@nju.edu.cn}

\input{macros}

\usepackage{algorithm}
\usepackage{algorithmic}

\begin{document}

\begin{abstract}
Running large language models on consumer devices such as laptops and desktops is challenging because model weights often exceed GPU memory capacity, making offloading inference necessary to extend effective model capacity with CPU memory. Existing offloading systems, however, typically rely on coarse layer-level or expert-level scheduling, which overlooks substantial heterogeneity among tensors within the same layer and adapts poorly to changing hardware load conditions on such devices. This paper presents \system{}, a hybrid CPU-GPU inference system for consumer devices that performs offloading at tensor granularity. \system{} combines static tensor placement with load-aware dynamic transfer, and introduces asynchronous CPU-GPU coordination to efficiently schedule hardware storage, data movement, and computation across heterogeneous backends. We implement \system{} and evaluate it on representative consumer platforms using both dense and MoE models. Compared with existing systems, \system{} improves prefill throughput by up to 1.94$\times$ and decode throughput by up to 3.29$\times$, while also increasing GPU utilization and making more effective use of PCIe bandwidth. These results show that \system{} can substantially improve the user experience of local LLM deployment on personal consumer devices.
\end{abstract}

\maketitle
\pagestyle{plain}

\input{chapters/1-introduction}
\input{chapters/2-background}
\input{chapters/3-motivation}
\input{chapters/4-system}
\input{chapters/5-evaluation}
\input{chapters/6-related-work}

\input{chapters/7-conclusion}

\bibliographystyle{ACM-Reference-Format}
\bibliography{paper}

\end{document}

%% file: macros.tex
\newcommand{\system}{\textsc{ATSInfer}}

%% file: chapters/1-introduction.tex
\section{Introduction}
\label{sec:introduction}

Large language models~(LLMs) have rapidly become a foundation for language understanding, code generation, question answering, and interactive assistants~\cite{gpt,qwen3,deepseek,llama4}. As demand grows for privacy, offline availability, and predictable latency, deployment is increasingly moving from cloud servers to consumer devices such as laptops and desktops. This shift changes the optimization target. Unlike server-side LLM serving, which often relies on many concurrent requests and large batches to maximize throughput, local deployment usually runs at low concurrency, making per-request responsiveness the most important performance metric.

\begin{figure}[t]
	\centering
	\includegraphics[width=\linewidth]{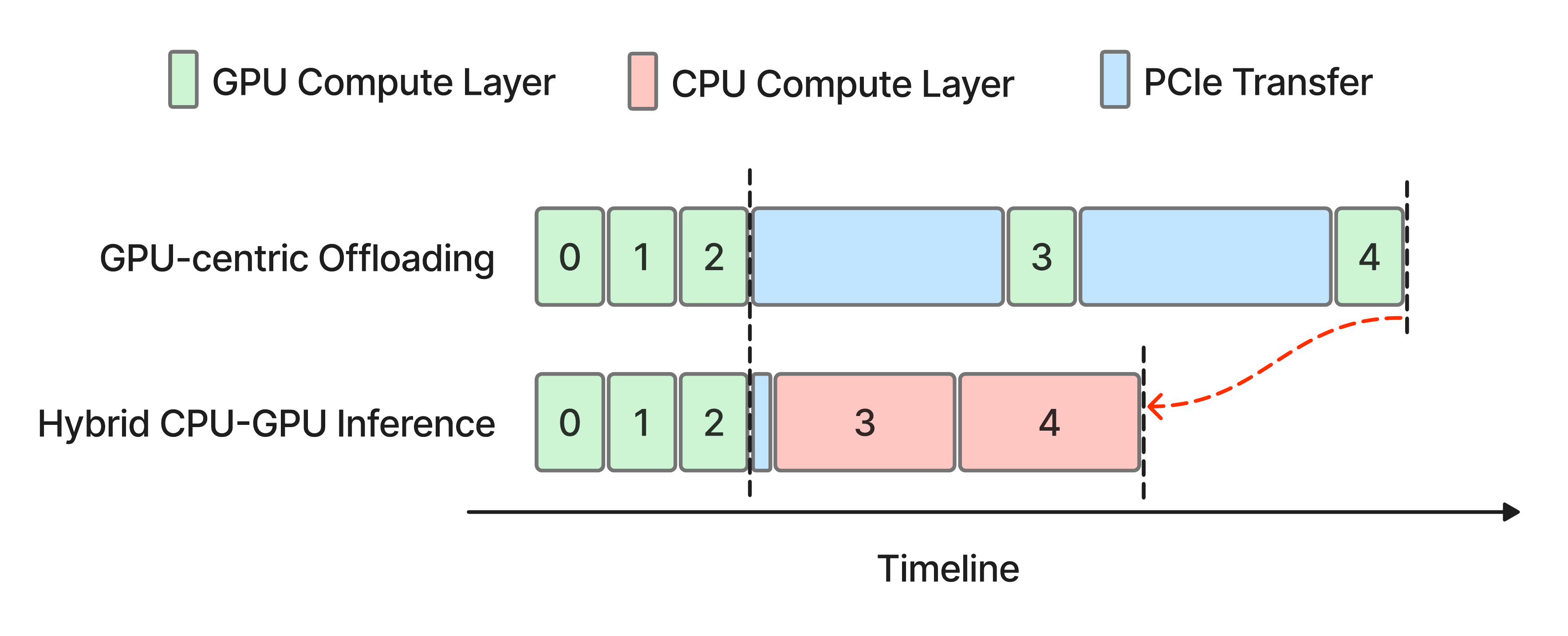}
	\caption{Execution timelines for GPU-centric offloading and hybrid CPU-GPU inference on consumer devices. Hybrid inference replaces long PCIe transfers of model weights with shorter transfers of activations, but CPU execution remains substantially slower than GPU execution and becomes the new bottleneck.}
	\label{fig:intro:bottleneck-shift}
	\Description{A timeline comparison between GPU-centric offloading and hybrid CPU-GPU inference on consumer devices. The GPU-centric design spends substantial time transferring model weights over PCIe. The hybrid design reduces transfer time by moving activations instead, but CPU execution remains much longer than GPU execution and becomes the dominant bottleneck.}
\end{figure}

Deploying modern LLMs on consumer devices remains difficult because model weights often exceed GPU memory capacity even after quantization. Offloading~\cite{flexgen} is therefore unavoidable for practically useful models. A common approach is GPU-centric offloading, in which the GPU remains the primary compute device and model weights are moved across a storage hierarchy on demand~\cite{flexgen,huggingface,deepspeed,vllm,sglang}. However, this design is fundamentally constrained by the bandwidth mismatch between GPU memory and PCIe: consumer GPUs typically provide on-device memory bandwidth in the hundreds to over a thousand GB/s, whereas PCIe 4.0 provides at most 32~GB/s of host-device bandwidth. Once weights must be fetched repeatedly from host memory, PCIe transfer becomes the bottleneck and the GPU is difficult to keep fully utilized~\cite{powerinfer}.

This bottleneck has motivated hybrid CPU-GPU inference, in which the CPU participates directly in model computation instead of treating it only as a source of transferred weights~\cite{llama.cpp,flexinfer,powerinfer,llminaflash,ktransformers,hybrimoe,moelightning,dali,promoe,moe-infinity,adapmoe}. Hybrid execution reduces the need to move every tensor to the GPU and expands the design space beyond transfer-computation overlap alone. However, it introduces a new bottleneck: CPU execution can dominate end-to-end latency, especially on consumer devices where CPU and GPU performance is highly asymmetric on massively parallel operators such as matrix multiplication~\cite{ktransformers,powerinfer}. Once a substantial fraction of the model remains on the CPU, long CPU execution leaves the GPU idle for extended periods and sharply reduces the benefit of offloading~\cite{llama.cpp,flexinfer}. Figure~\ref{fig:intro:bottleneck-shift} illustrates how the primary bottleneck shifts from PCIe transfer to CPU execution.

To relieve CPU-side bottlenecks under tight GPU memory constraints, we focus on two limitations of existing hybrid inference systems.

\begin{itemize}

	\item \textbf{Coarse-grained placement.} Existing hybrid systems typically place weights at layer granularity~\cite{llama.cpp,flexinfer,flexgen,deepspeed}, or at expert granularity~\cite{ktransformers,hybrimoe,dali,fiddler,moelightning,promoe,moe-infinity,adapmoe} for Mixture-of-Experts~(MoE) models~\cite{gpt-oss,qwen3,qwen3.5,deepseek,mixtral,llama4}. This granularity leaves substantial intra-layer heterogeneity unexplored.
	
	\item \textbf{Load-unaware scheduling.} Existing runtime policies are either fixed~\cite{flexgen,deepspeed,flexinfer,ktransformers} or adapt only to input-dependent signals such as sequence length or expert activation~\cite{hybrimoe,moelightning,promoe,moe-infinity}. However, they do not respond to changing device conditions during execution.
	
\end{itemize}

To address these challenges, we present \system{}, a tensor-granularity hybrid CPU-GPU inference system with load-aware dynamic transfer. \system{} combines three complementary mechanisms: \emph{asynchronous CPU-GPU scheduling} to coordinate storage, data movement, and computation across heterogeneous backends; \emph{static tensor placement} to determine GPU residency under memory and switching-cost constraints; and \emph{load-aware dynamic transfer} to adjust runtime tensor movement according to inference phase and backend load.

Together, these three components improve hybrid CPU-GPU inference for local low-concurrency workloads, where trade-offs among residency, transfer, and execution directly determine latency. In summary, this paper makes the following contributions:

\begin{itemize}
	\item We empirically analyze hybrid CPU-GPU inference on consumer devices and show why coarse layer- or expert-level placement often fails to minimize latency.

	\item We design \system{} around three coordinated mechanisms: asynchronous CPU-GPU scheduling, static tensor placement, and load-aware dynamic transfer, which together substantially alleviate the CPU bottleneck in hybrid inference and improve end-to-end execution efficiency.

	\item We implement \system{} by extending llama.cpp with about 15{,}000 lines of C++ code to support tensor-level offloading for both dense and MoE models.

	\item Compared with llama.cpp under the same GPU VRAM budget, \system{} improves prefill throughput by up to 1.94$\times$, improves decode throughput by up to 3.29$\times$, and increases average GPU utilization during decode by about 70\%.
\end{itemize}

%% file: chapters/2-background.tex
\section{Background}
\label{sec:background}

\subsection{CPU and GPU Architecture for Hybrid Inference}
\label{sec:bg:architecture}

\begin{figure}[t]
    \centering
    \includegraphics[width=0.95\linewidth]{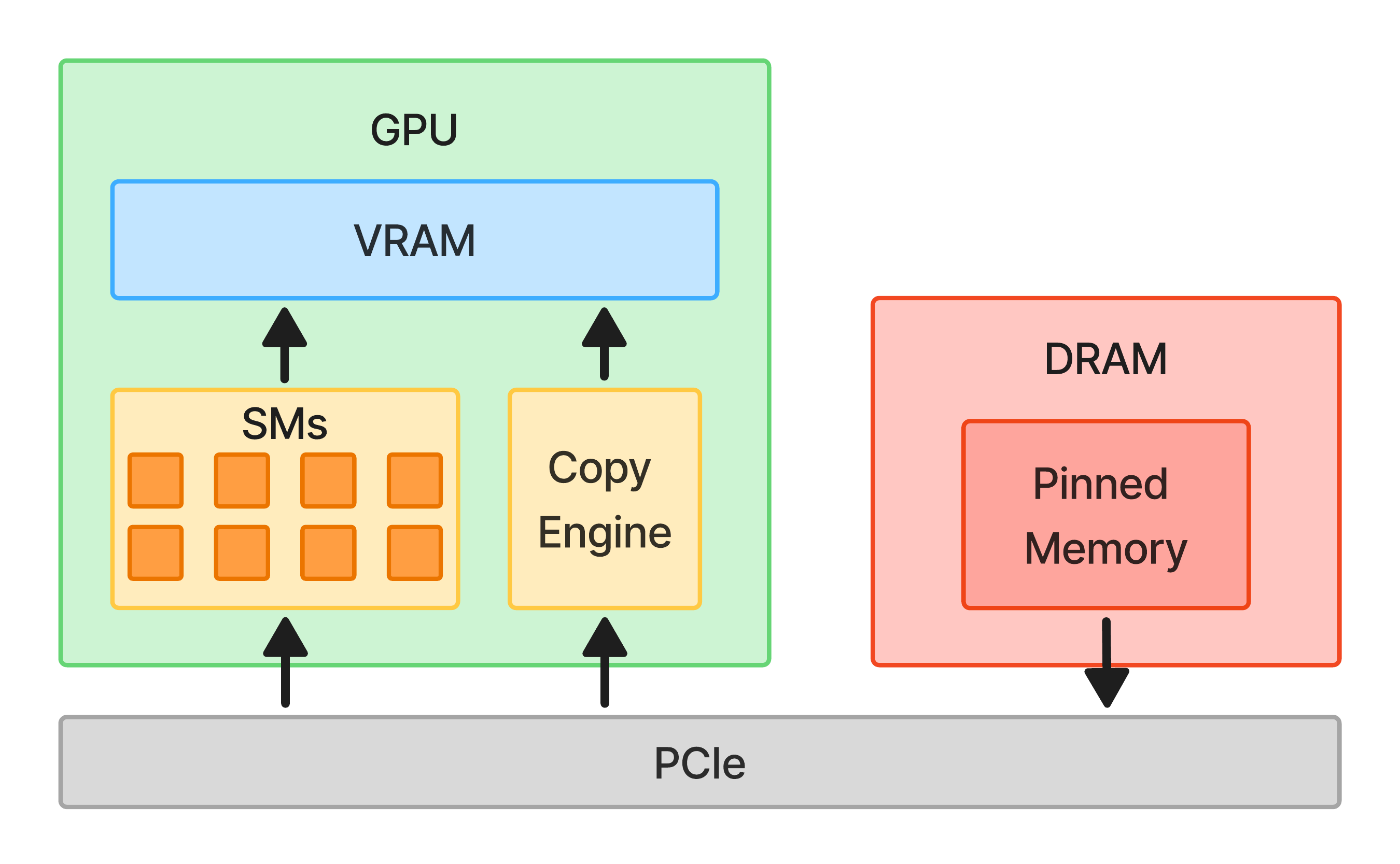}
    \caption{Two host-to-device transfer paths on NVIDIA GPUs. Besides the hardware Copy Engine path, mapped and pinned host memory also allows an SM-driven copy kernel to fetch data through Zero-Copy.}
    \label{fig:bg:transfer-paths}
    \Description{A diagram of two host-to-device transfer paths on an NVIDIA GPU. One path uses the hardware Copy Engine to move data from host memory to GPU memory over PCIe. The other path uses an SM-driven copy kernel with mapped and pinned host memory through Zero-Copy. The figure shows that the SM-driven path can run concurrently with Copy Engine transfers while sharing PCIe bandwidth.}
\end{figure}

Hybrid inference executes different parts of an LLM across the CPU and GPU. Its efficiency depends on several architectural properties of the two processors.

\noindent\textbf{CPU Instruction Set Optimizations.}
Modern x86 CPUs provide Single Instruction, Multiple Data~(SIMD) extensions for accelerating matrix multiplication, the core operation in LLM inference. AVX2 and AVX-512~\cite{avx512} support wide-vector execution on mainstream CPUs, while Advanced Matrix Extensions~(AMX) offer higher throughput but are largely limited to server-class processors such as Intel Xeon~\cite{amx}. This distinction matters in practice: llama.cpp's ggml backend emphasizes broad SIMD support, whereas systems such as KTransformers rely on AMX-optimized kernels that are typically unavailable on consumer CPUs~\cite{intel}.

\noindent\textbf{GPU Concurrent Execution and Data Transfer.}
Overlapping data transfers with GPU computation is a standard way to improve concurrency. NVIDIA GPUs support CUDA streams, where operations within a stream execute in order but operations across streams may overlap~\cite{cuda}. This allows systems to pipeline host-device transfers with GPU kernels. However, concurrent execution of multiple transfers is more constrained on consumer devices.

On NVIDIA GPUs, host-device transfers are handled primarily by Copy Engines, which move data over PCIe without consuming Streaming Multiprocessor~(SM) cycles. Consumer-grade RTX GPUs typically expose only one Copy Engine, so even transfer requests issued from different streams execute sequentially on that engine. To supplement this hardware path, systems can use a \emph{copy kernel}. When host memory is mapped and pinned, NVIDIA's Zero-Copy feature allows GPU threads to access it directly, enabling SM-driven transfers to overlap with Copy Engine-driven transfers while sharing PCIe bandwidth, as illustrated in Figure~\ref{fig:bg:transfer-paths}. This distinction is important for hybrid inference as it prevents high-priority transfers from being blocked by long-running transfer tasks.

\subsection{LLM Inference on Consumer Devices}
LLM inference on consumer devices differs from datacenter deployment in three key respects.

First, local deployment faces tighter resource constraints. Server deployments can often keep the full model on one or more GPUs, whereas consumer devices typically cannot fit even quantized models entirely in VRAM~\cite{vllm,sglang}. As a result, offloading becomes a common choice for LLM inference on consumer devices.

Second, the role of the key-value cache~(KV-cache) differs from that in server deployments. Servers optimize for throughput under high concurrency and therefore maintain large aggregate KV-caches across many requests, making mechanisms such as paging~\cite{vllm}, prefix sharing~\cite{vllm,sglang}, and large KV-cache pools~\cite{mooncake} important for reducing serving cost and sustaining throughput. In local interactive inference, the KV-cache is typically far smaller than the model weights, often occupying only hundreds of MB rather than tens of GB. Thus, it is usually kept with the backend executing the corresponding attention operator to avoid extra transfers in practice.

Third, local deployment must handle both \emph{prefill} and \emph{decode}~\cite{vllm} on the same machine. Prefill processes the full prompt and is more compute-intensive, whereas decode generates one token per step and is more memory-intensive~\cite{splitwise}. Server systems often separate these phases by prefill-decode disaggregation technique~\cite{splitwise,mooncake} and optimize them independently, but local systems must handle both under the same hardware constraints. 

%% file: chapters/3-motivation.tex
\section{Motivation}
\label{sec:motivation}

This section provides empirical evidence for the two design pressures highlighted in the introduction: coarse offloading granularity and load-unaware scheduling on consumer hardware.

\subsection{Limits of Coarse Offloading Granularity}

\begin{figure}[t]
\centering
\includegraphics[width=\linewidth]{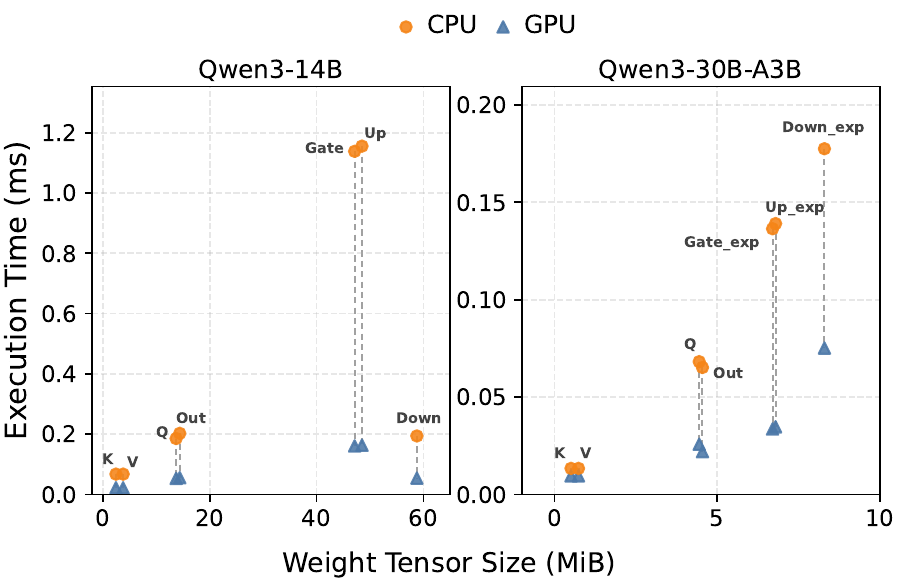}
\caption{Sizes and backend execution times of major weight tensors within one layer for Qwen3-14B and Qwen3-30B-A3B on an RTX 3060 + Intel i7-11800H laptop. For MoE operators, the reported weight size includes only the activated experts rather than all experts.}
\label{fig:motivation:tensor-heterogeneity}
\Description{A figure comparing the sizes and backend execution times of major weight tensors within one layer for the Qwen3-14B and Qwen3-30B-A3B models on an RTX 3060 and Intel i7-11800H laptop. The figure shows that tensors in the same layer differ substantially in offloading benefit, with some tensors offering much larger speedups from GPU execution than others.}
\end{figure}

Existing offloading systems typically make placement decisions at a coarse granularity. For dense Transformer models, the common unit is the layer, while for MoE models some systems refine the unit to experts.

However, layer- or expert-level placement remains too coarse for hybrid inference on consumer devices. A layer or expert comprises multiple weight tensors, and the operators consuming these tensors differ substantially in compute intensity, memory access patterns, and kernel implementation efficiency. As a result, the latency benefit of GPU residency is not uniform: on consumer hardware, moving one tensor to the GPU may yield a much larger speedup than moving another tensor of comparable size.

This heterogeneity exposes an optimization space that coarse-grained policies cannot exploit. Figure~\ref{fig:motivation:tensor-heterogeneity} illustrates this point by comparing the sizes and backend execution times of major weight tensors within one layer during decode. Under tight VRAM budgets, offloading decisions should instead prioritize tensors by the latency reduction they deliver per byte of GPU memory. 

\subsection{Limits of Load-unaware Scheduling}

Many systems rely on policies determined before execution and keep them unchanged at runtime. For example, llama.cpp~\cite{llama.cpp} offloads selected compute-heavy operators to the GPU when batch size or sequence length exceeds a hard-coded threshold, while KTransformers uses Expert Deferral to overlap delayed CPU-side expert computation with later GPU work during decode~\cite{ktransformers}.

Some MoE systems do adapt at runtime, but mainly to input-dependent behavior rather than device conditions. Specifically, they adjust expert retention and replacement based on activation patterns to improve GPU hit rates and reduce CPU load~\cite{hybrimoe,moelightning,promoe,moe-infinity,adapmoe}. Even so, these mechanisms still respond primarily to model structure, request-level demand, and expert activation, rather than to transient hardware bottlenecks.

Crucially, these approaches overlook device-level performance variability on consumer hardware. Unlike server deployments, which usually run in stable and dedicated environments, local inference often co-runs with browsers, IDEs, media applications, and other background workloads. These applications dynamically change the CPU, GPU, and memory resources available to inference. In addition, sustained high utilization on laptops and mobile devices can trigger thermal throttling or power limits, reducing processor frequencies over time. As a result, the relative benefits of CPU execution, GPU execution, and host-device transfer can shift substantially during execution, making both static policies and input-driven runtime policies suboptimal.

\begin{figure}[t]
\centering
\includegraphics[width=\linewidth]{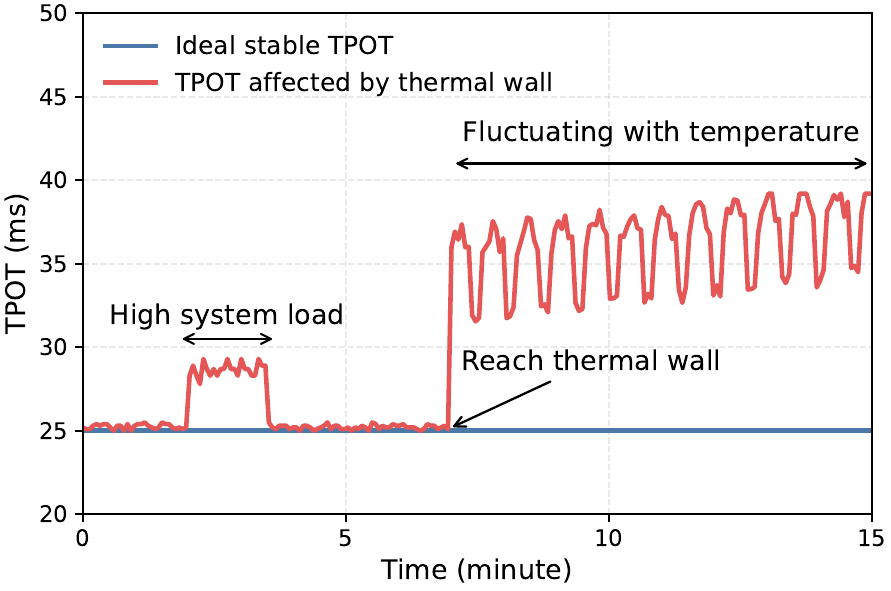}
\caption{TPOT over time under dynamic local conditions. The figure highlights two realistic sources of performance variation on consumer devices: interference from background tasks and later degradation caused by a thermal wall.}
\label{fig:motivation:load-variation}
\Description{A TPOT-over-time figure showing that token latency first increases under background-task interference and later changes again when the device reaches a thermal wall, illustrating dynamic performance variation during local inference.}
\end{figure}

Figure~\ref{fig:motivation:load-variation} illustrates this mismatch by showing how Time per output token~(TPOT) evolves over time as the system first experiences interference from background tasks and then hits a thermal wall. Therefore, practical local inference requires a runtime scheduler that reacts to transient bottlenecks. If CPU throughput degrades due to contention or thermal throttling, the system should shift profitable computation to the GPU. Conversely, if GPU performance or PCIe transfer efficiency becomes the limiting factor, it should reduce transfers and rely more on CPU execution. The key implication is that runtime transfer decisions must be load-aware rather than fixed or merely input-driven.

Taken together, these observations motivate the design of \system{}, a high-performance local inference system for consumer devices that supports tensor-level fine-grained offloading and load-aware dynamic transfer.

%% file: chapters/4-system.tex
\section{\system{} Design}
\label{sec:system}

\subsection{Overview of \system{}}

\begin{figure}[t]
\centering
\includegraphics[width=0.95\linewidth]{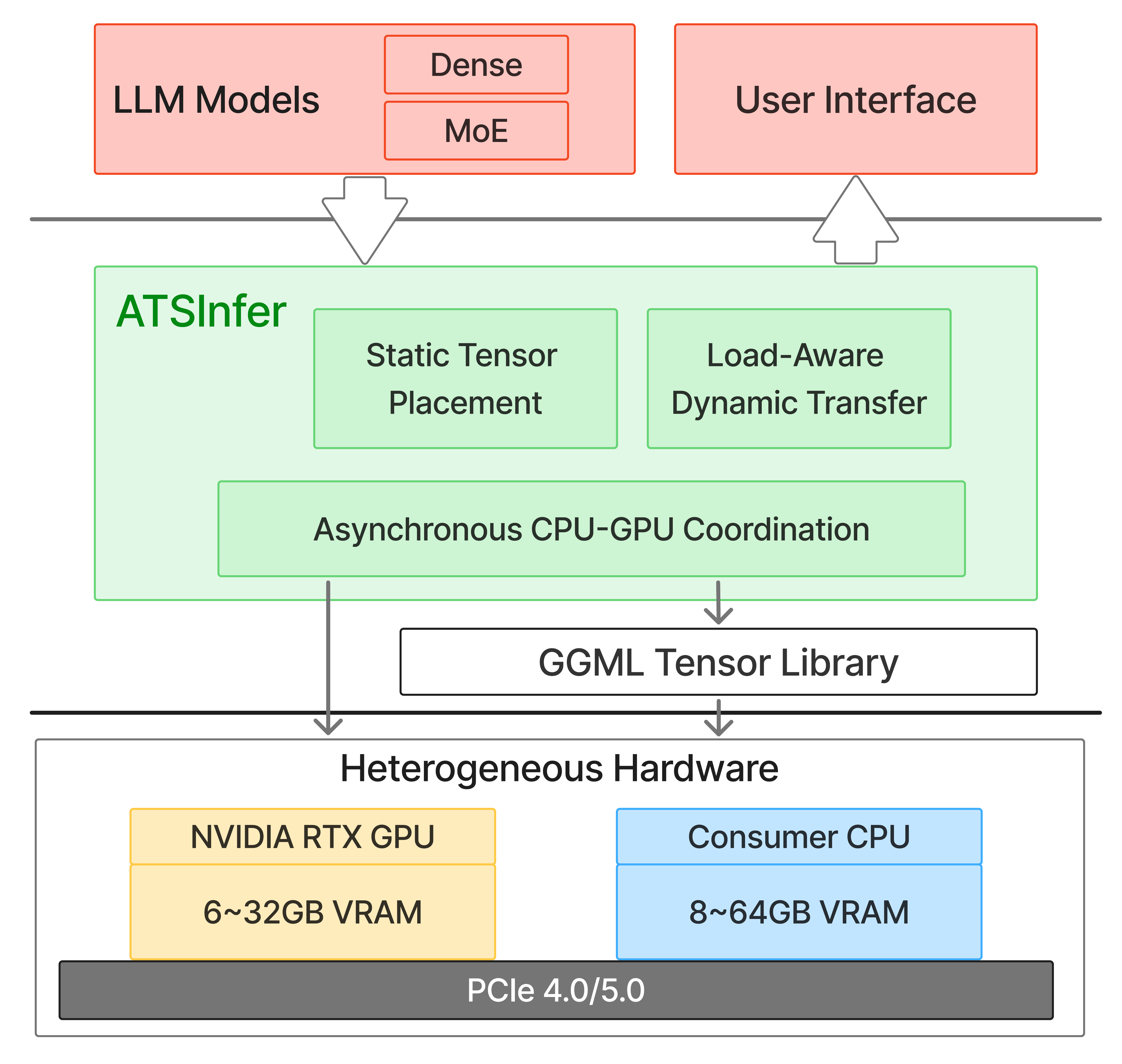}
\caption{System architecture of \system{}.}
\label{fig:system:architecture}
\Description{An architecture diagram of \system{} showing asynchronous CPU-GPU coordination as the execution substrate, together with static tensor placement and load-aware dynamic transfer for tensor-granularity hybrid inference.}
\end{figure}

Figure~\ref{fig:system:architecture} shows the overall architecture of \system{}. Built on top of llama.cpp~\cite{llama.cpp}, \system{} adds a tensor-granularity execution path organized around three components: an asynchronous CPU-GPU coordination substrate, a static tensor placement mechanism, and a load-aware dynamic transfer mechanism. Together, these components directly orchestrate CPU and GPU storage, computation, and data movement across heterogeneous hardware, while exposing user-friendly interfaces and preserving llama.cpp's broad support for dense and MoE models.

\begin{figure*}[t]
\centering
\includegraphics[width=0.95\textwidth]{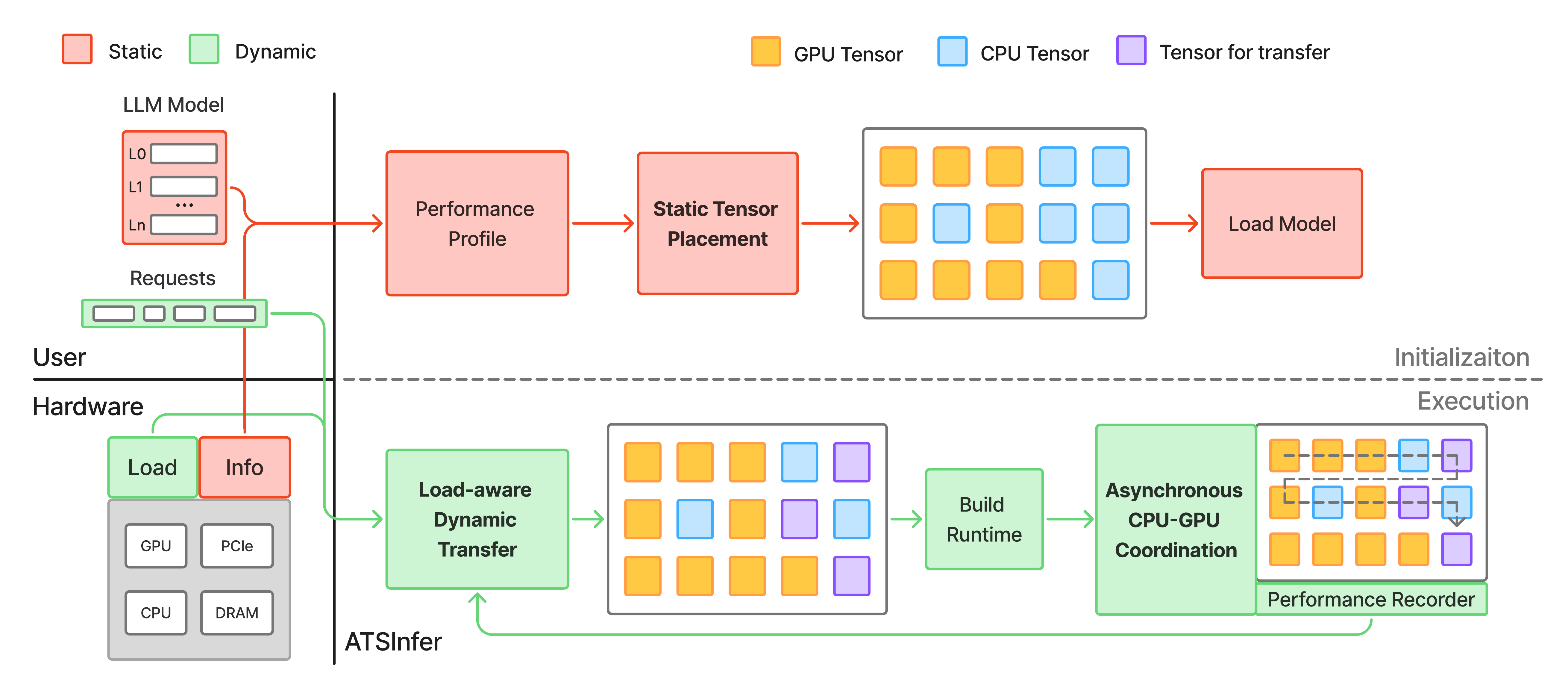}
\caption{Workflow overview of \system{}.}
\label{fig:system:workflow}
\Description{An operation-flow diagram of \system{} showing offline profiling and initialization, followed by online scheduling before prefill and decode, execution-plan construction, and overlapped CPU-GPU execution.}
\end{figure*}

Figure~\ref{fig:system:workflow} summarizes how these components interact during execution. The workflow consists of an offline initialization phase and an online execution phase.

In the offline phase, \system{} collects hardware characteristics and profiles tensor-level execution behavior on the target backends. Using these measurements, it derives empirical performance estimates for CPU execution, GPU execution, and transfer, then computes a default tensor placement plan under the available memory budget. It also initializes the memory layout required for hybrid inference, including the GPU resident region, CPU pinned-memory storage, and the temporary GPU buffers used for dynamic transfer.

In the online phase, \system{} serves each request through repeated scheduling and execution rounds. Before prefill and each decode step, it uses recent measurements to estimate current CPU speed, GPU speed, transfer cost, and overlap opportunity. It then selects a subset of CPU-resident tensors for temporary promotion to the GPU. Next, it allocates space for these tensors in the GPU temporary buffers, constructs an execution plan with backend assignments, transfer events, and synchronization points, and executes the round while recording the performance information needed for subsequent scheduling decisions.

\subsection{CPU-GPU Coordination}

To enable overlap between the two kinds of data movement and computation, \system{} first organizes CPU memory and GPU memory carefully. On top of this memory layout, it builds a pipelined CPU-GPU coordination mechanism designed to maximize concurrency while minimizing unnecessary synchronization.

\subsubsection{Memory Layout}

\begin{figure}[t]
\centering
\includegraphics[width=\linewidth]{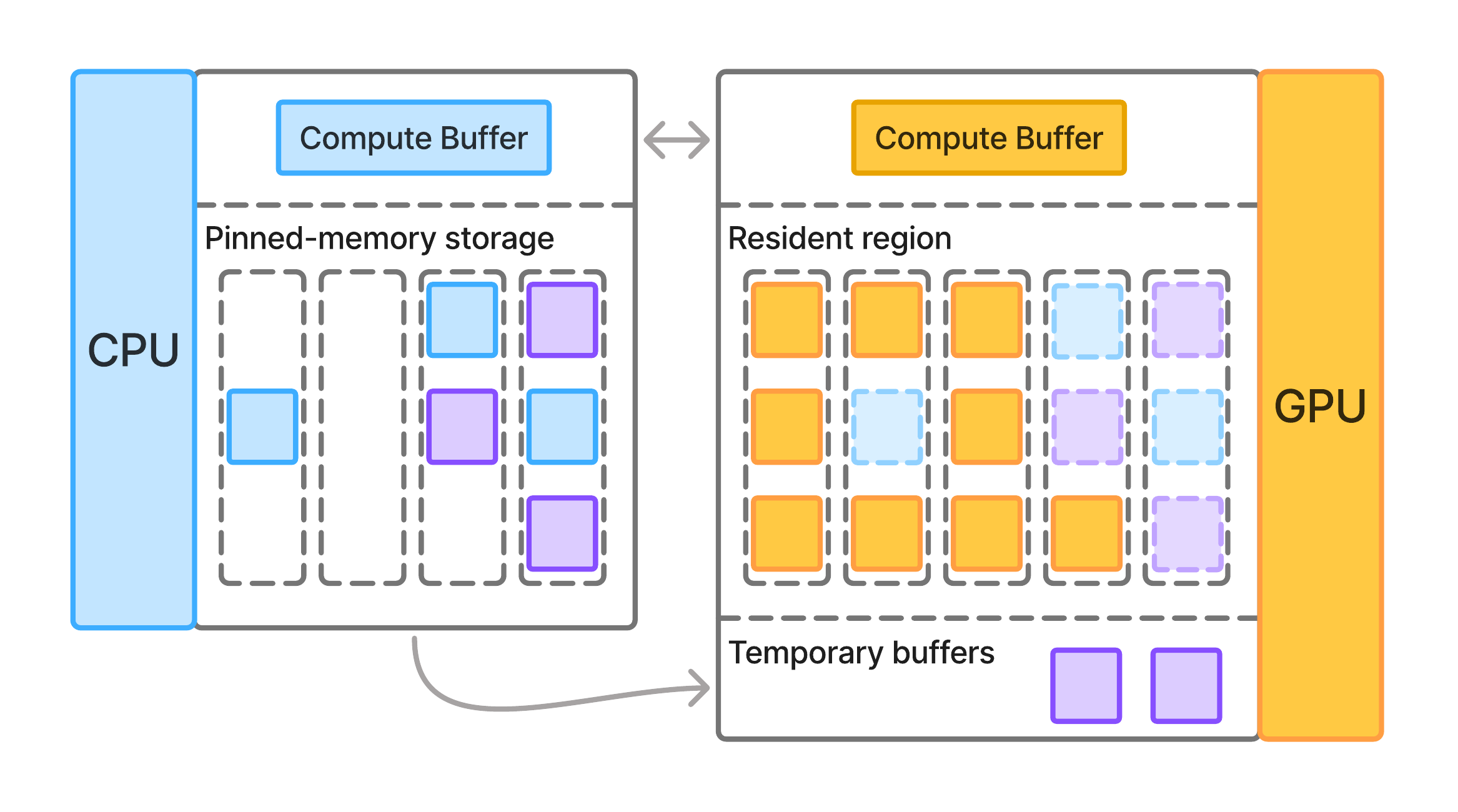}
\caption{Memory layout of \system{}.}
\label{fig:system:memory-layout}
\Description{A memory-layout diagram of \system{} showing the GPU resident region, CPU pinned-memory storage, GPU temporary buffers, and the compute buffers managed by the ggml backend.}
\end{figure}

As shown in Figure~\ref{fig:system:memory-layout}, \system{} organizes model weights into three logical memory regions to support fine-grained hybrid execution.

The first is the \emph{GPU resident region}, which stores tensors selected by static placement. These tensors are loaded into GPU memory before inference and kept resident whenever possible, so frequently used high-value operators can execute without repeated transfer overhead.

The second is \emph{CPU pinned-memory storage}, which holds tensors that cannot remain permanently on the GPU because of memory limits. Keeping these tensors in mapped and pinned memory improves the stability and efficiency of CPU-to-GPU transfer and makes asynchronous movement with copy kernels practical at runtime.

The third is a set of \emph{GPU temporary buffers} for tensors that are promoted from CPU memory to the GPU on demand. Rather than allocating a separate GPU region for every transferred tensor, \system{} organizes temporary storage by tensor lifetime and reuses the same buffer space across tensors whose live ranges do not overlap. In practice, tensors that play the same role in different layers can often share the same temporary space because they are consumed sequentially. This reuse strategy reduces peak GPU memory consumption without introducing additional synchronization complexity.

This layout cleanly separates long-lived residency from short-lived runtime promotion. In addition to these three regions, both the CPU and the GPU maintain their own compute buffers for operator execution; these buffers are preallocated and efficiently managed by the underlying ggml backend.

\subsubsection{Asynchronous CPU-GPU Scheduling}

\begin{figure}[t]
\centering
\includegraphics[width=\linewidth]{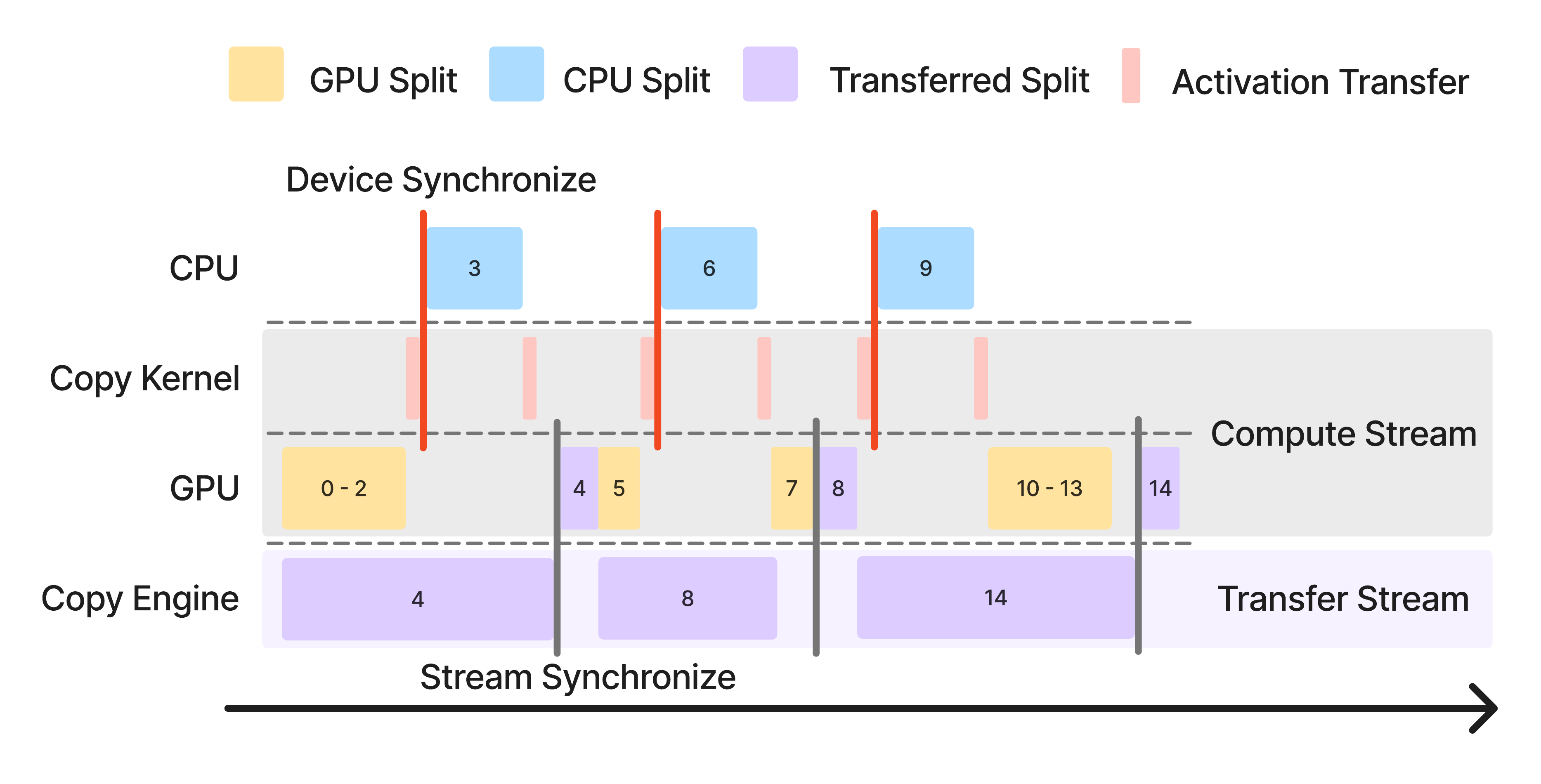}
\caption{Asynchronous CPU-GPU execution pipeline in \system{}.}
\label{fig:system:pipeline}
\Description{A pipeline diagram of \system{} showing compute on CPU and GPU, cross-device activation transfer, asynchronous weight transfer, and synchronization among the compute stream and transfer stream.}
\end{figure}

Before prefill and each decode round, \system{} constructs an execution plan for the current round and reassigns GPU temporary-buffer space accordingly. To control synchronization precisely, it partitions the computation graph at two kinds of positions: backend-switch boundaries and operators that consume tensors stored in temporary GPU buffers. The resulting plan is a sequence of ordered splits with explicit transfer and synchronization boundaries. Figure~\ref{fig:system:pipeline} illustrates a representative CPU-GPU execution pipeline.

\system{} uses two streams to drive execution. The \emph{compute stream} serializes operator execution together with the necessary transfer of activations across backends, while the \emph{transfer stream} handles asynchronous CPU-to-GPU movement of dynamically selected weights. Separating the two streams is not sufficient by itself: the system also maps them to different execution engines. Activation transfer associated with computation is performed by copy kernels running on the GPU SMs, whereas weight movement is handled by the dedicated CUDA copy engine. Although the two transfer types still share PCIe bandwidth, this design prevents small activation transfers from being blocked behind large weight transfers and therefore preserves downstream computation overlap.

To maximize overlap, \system{} launches weight transfers as early as possible without violating data dependencies. In dense models, weight transfer can typically start as soon as the required temporary-buffer slot becomes available. In MoE models, expert-weight transfer must wait until routing has been resolved, so that only the experts selected by the current token are moved to the GPU. Although MoE models may offer only a limited overlap window between computation and transfer, decode typically activates only a small number of experts, which keeps the corresponding transfer time short.

Correctness relies on explicit synchronization among the CPU thread, the compute stream, and the transfer stream. When a GPU split finishes and triggers activation transfer, both operations remain ordered within the compute stream. When execution moves from GPU to CPU, the runtime synchronizes with the compute stream to ensure that the required data movement has completed. When asynchronous weight transfer starts or completes, lightweight GPU events coordinate cross-stream dependencies without forcing unnecessary global synchronization. Under these rules, \system{} overlaps transfer and computation while preserving execution consistency and avoiding races or premature temporary-buffer reuse.

\subsection{Static Tensor Placement}

Static placement determines which tensors should remain resident on the GPU before inference starts. As discussed in Section~\ref{sec:motivation}, the benefit of offloading varies substantially across tensors, making tensor-granularity placement essential.

\subsubsection{Empirical Performance Density}

To decide which tensors deserve permanent GPU residency, \system{} relies on measured execution behavior rather than a purely analytic model. In practice, tensor performance depends on multiple hardware- and implementation-specific factors, so theoretical metrics alone are often insufficient for guiding placement. For tensor $i$, let $s_i$ denote its memory footprint, and let $t_i^c$ and $t_i^g$ denote the measured execution time of the corresponding operator on CPU and GPU. We define the \emph{empirical performance density} on backend $b \in \{c,g\}$ as 
\[
k_i^b = \frac{t_i^b}{s_i}.
\]

This metric measures the execution cost associated with each unit of model memory, enabling comparison across tensors of different sizes.

Compared with traditional compute-intensity metrics, empirical performance density has three advantages:
\begin{itemize}
	\item It reflects observed runtime behavior, because $t_i$ is derived from direct measurement and therefore captures both hardware characteristics and kernel implementation efficiency.
	\item It jointly captures compute and memory effects, because the measured latency already incorporates both computational bottlenecks and memory-access overheads.
	\item It is directly actionable for system optimization, because it maps naturally to the trade-off between latency benefit and GPU memory consumption.
\end{itemize}

\begin{figure}[t]
\centering
\includegraphics[width=\linewidth]{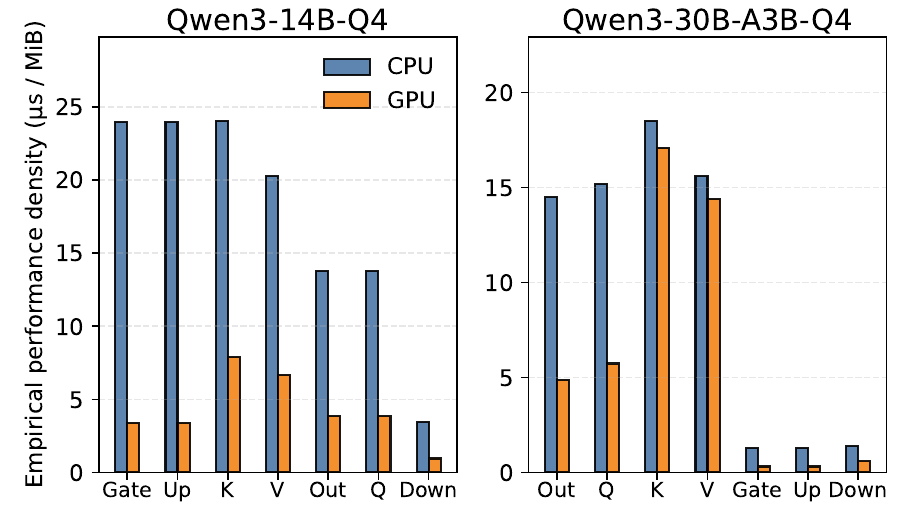}
\caption{Empirical performance density of representative tensors during decode, sorted by the density difference between CPU and GPU execution. For MoE expert tensors, the size is computed as the total size of all expert weights of the operator, because the actual routed experts are unknown before model loading and all experts within the same operator follow a unified placement decision.}
\label{fig:system:epd}
\Description{A comparison of empirical performance density across representative tensors during decode, showing substantial variation within the same layer.}
\end{figure}

As discussed in Section~\ref{sec:motivation}, tensor-level heterogeneity is substantial enough that coarse layer- or expert-level placement fails to capture the full performance opportunity. Figure~\ref{fig:system:epd} shows that empirical performance density varies substantially across tensors and backends. This observation motivates tensor-granularity placement and suggests that uniform treatment of tensors within a layer or expert can lead to suboptimal resource utilization.

\subsubsection{Problem Formulation}

Using the tensor-level quantities defined above, including $s_i$, $t_i^c$, $t_i^g$, and $r_i$, we now formulate static placement. Let $b_i \in \{\mathrm{CPU}, \mathrm{GPU}\}$ denote the backend assigned to tensor $i$, and let $M$ denote the GPU memory budget available for static residency.

When two adjacent tensors are assigned to different backends, the system incurs an additional switching cost due to activation transfer. Let $c_i$ denote the switching cost at the boundary between tensors $i-1$ and $i$. If the measured PCIe bandwidth is $B_{\mathrm{pcie}}$ and the total size of the transferred inputs at this boundary is $S_{\mathrm{in},i}$, we estimate$c_i = S_{\mathrm{in},i} / B_{\mathrm{pcie}}$.

Static placement therefore solves the following optimization problem:
\[
\max \left( \sum_{i=1}^{n} r_i\,\mathbf{1}\{b_i = \mathrm{GPU}\} - \sum_{i=2}^{n} c_i\,\mathbf{1}\{b_i \neq b_{i-1}\} \right)
\]
subject to
\[
\sum_{i=1}^{n} s_i\,\mathbf{1}\{b_i = \mathrm{GPU}\} \le M, \qquad b_i \in \{\mathrm{CPU},\mathrm{GPU}\}.
\]
The first term captures the latency reduction from GPU-resident tensors, while the second penalizes backend fragmentation. Overall, the problem is a knapsack-style optimization with an additional penalty on adjacent backend switches. 

\subsubsection{Tensor Placement Solver}

Before solving the optimization, \system{} first profiles the target platform to obtain all required inputs. It measures per-operator latency on CPU and GPU to estimate $t_i^c$, $t_i^g$, and thus $r_i$; records tensor sizes to obtain $s_i$; and estimates switching costs $c_i$ from activation sizes and measured PCIe bandwidth. The resulting static placement is therefore specialized to the target model, runtime, and hardware.

For dense models, \system{} directly solves the knapsack-style optimization above with a standard dynamic program and then recovers the placement by backtracking. For MoE models, \system{} integrates the solver with an expert/non-expert partitioning policy to avoid overestimating the value of expert residency.

\begin{algorithm}[t]
\caption{Static Tensor Placement}
\label{alg:system:static-placement}
\begin{algorithmic}[1]
\STATE \textbf{Input:}
\STATE \hspace{1.5em}model $\mathcal{M}$;
\STATE \hspace{1.5em}GPU memory budget $M$
\STATE \textbf{Output:}
\STATE \hspace{1.5em}placement decision $b$

\STATE $s \gets \textsc{ReadTensorSizes}(\mathcal{M})$;
\STATE $G \gets \textsc{BuildComputationGraph}(\mathcal{M})$;
\STATE $(t^c, t^g, c) \gets \textsc{Profile}(G)$;
\STATE $r \gets t^c - t^g$;

\IF{$\mathcal{M}$ is dense}
	\STATE $b \gets \textsc{SolveKnapsackDP}(G, s, r, c, M)$;
\ELSE
	\STATE $(T_{\mathrm{exp}}, T_{\mathrm{nonexp}}) \gets \textsc{PartitionTensors}(G)$;
	\STATE $s_{\mathrm{nonexp}} \gets \textsc{TotalSize}(T_{\mathrm{nonexp}}, s)$;
	\IF{$M \ge s_{\mathrm{nonexp}}$}
		\STATE $b_{\mathrm{nonexp}} \gets \mathrm{GPU}$;
		\STATE $b_{\mathrm{exp}} \gets \textsc{SolveKnapsackDP}(T_{\mathrm{exp}}, s, r, c, M - s_{\mathrm{nonexp}})$;
	\ELSE
		\STATE $b_{\mathrm{exp}} \gets \mathrm{CPU}$;
		\STATE $b_{\mathrm{nonexp}} \gets \textsc{SolveKnapsackDP}(T_{\mathrm{nonexp}}, s, r, c, M)$;
	\ENDIF
\ENDIF
\STATE \textbf{Return} $b$
\end{algorithmic}
\end{algorithm}

If the GPU memory budget $M$ is discretized as the dynamic-programming memory dimension, the dense-model solver runs in $O(nM)$ time and uses $O(nM)$ space. For MoE models, tensor partitioning and capacity checks add only linear preprocessing overhead, so overall complexity is still dominated by the dynamic program. In practice, \system{} quantizes both $s_i$ and $M$ at MB granularity. This discretization preserves sufficient accuracy for placement while substantially shrinking the search space and reducing runtime and memory overhead. Because static placement is performed offline or during initialization, its cost is amortized across subsequent inference requests.

\subsection{Load-Aware Dynamic Transfer}

Static placement provides an effective default configuration, but it is insufficient once inference begins. To further relieve the CPU bottleneck, \system{} selectively transfers some CPU-resident tensors to the GPU for execution and overlaps their weight movement with preceding computation, so that only the exposed portion of transfer time remains on the critical path. This design exploits otherwise idle PCIe bandwidth to shift part of the workload away from the CPU, thereby combining GPU-centric offloading with hybrid CPU-GPU inference and balancing pressure across CPU, GPU, and PCIe. Because its benefit depends on current CPU speed, GPU speed, transfer bandwidth, and overlap opportunity, \system{} makes these transfer decisions online in a load-aware manner rather than fixing them entirely through offline profiling.

\begin{algorithm}[t]
\caption{Dynamic Transfer Scheduling}
\label{alg:system:dp-transfer}
\begin{algorithmic}[1]
\STATE \textbf{Input:}
\STATE \hspace{1.5em}default placement $b$;
\STATE \hspace{1.5em}GPU/CPU execution times $t^g, t^c$;
\STATE \hspace{1.5em}activation-transfer costs $c$;
\STATE \hspace{1.5em}weight-transfer times $w$
\STATE \textbf{Output:}
\STATE \hspace{1.5em}runtime backend assignment $rb$

\STATE $seg[j,i] \gets \sum_{k=j+1}^{i-1} t_k^{b_k} + \sum_{k=j+1}^{i-1} c_k\,\mathbf{1}\{b_{k-1} \neq b_k\}$;
\STATE Initialize dynamic-programming states;

\FOR{$i=1$ to $n$}
	\IF{$b_i = \mathrm{GPU}$}
		\STATE Update $dp[i,\mathrm{GPU}]$ by extending $dp[i-1]$;
		\STATE Mark $dp[i,\mathrm{CPU}]$ as infeasible;
	\ELSE
		\STATE Update $dp[i,\mathrm{CPU}]$ by extending $dp[i-1]$;
		\STATE $best \gets \min\limits_{\substack{j < i \\ b_j = \mathrm{CPU}}} \left(dp[j,\mathrm{CPU}] + \max\bigl(w_i, seg[j,i]\bigr)\right)$;
		\STATE $dp[i,\mathrm{GPU}] \gets best + c_i\,\mathbf{1}\{b_{i-1} = \mathrm{GPU}\} + t_i^g$;
	\ENDIF
\ENDFOR
\STATE $(T^\star, rb^\star) \gets \arg\min_{rb \in \{\mathrm{GPU},\mathrm{CPU}\}} dp[n,rb]$;
\STATE Recover $rb_1,\dots,rb_n$ by backtracking;
\STATE \textbf{Return} $rb$
\end{algorithmic}
\end{algorithm}

\subsubsection{Problem Formulation}

Let $b=(b_1,\dots,b_n)$ denote the default backend placement produced by static placement, and let $rb_i \in \{\mathrm{CPU}, \mathrm{GPU}\}$ denote the runtime backend assignment in the current round. Tensors with $b_i=\mathrm{GPU}$ remain on the GPU, whereas tensors with $b_i=\mathrm{CPU}$ may be temporarily transferred to the GPU and executed there.

Let $t_i^c$, $t_i^g$, and $c_i$ denote the current CPU execution time, GPU execution time, and activation-transfer cost, respectively, and let $w_i$ denote the weight-transfer time of tensor $i$. Unlike static placement, these quantities are refreshed from recent runtime measurements before each scheduling round. For MoE expert weights in a decode round, $w_i$ is defined as the total transfer time of the activated experts rather than that of all experts in the operator. The objective is to minimize the end-to-end latency of the current round.

For a tensor that is CPU-resident by default but promoted to the GPU, only the non-overlapped fraction of transfer time contributes to the critical path. Let
\[
seg(j,i) = \sum_{k=j+1}^{i-1} t_k^{b_k} + \sum_{k=j+1}^{i-1} c_k\,\mathbf{1}\{b_{k-1} \neq b_k\}
\]
denote the amount of computation and activation-transfer time along the default path between the previous CPU-side endpoint $j$ and the start of tensor $i$, i.e., the overlap window available to hide weight transfer. The exposed transfer time is therefore
\[
\delta_i = \max\bigl(w_i - seg(j,i),\; 0\bigr).
\]

Under these definitions, total latency can be written as
\[
T(rb_1,\dots,rb_n)=\sum_{i=1}^n t_i^{rb_i}+\sum_{i=2}^n c_i\,\mathbf{1}\{rb_{i-1}\neq rb_i\}+\sum_{i\in \mathcal{G}} \delta_i,
\]
where $\mathcal{G}=\{i \mid b_i=\mathrm{CPU},\; rb_i=\mathrm{GPU}\}$ is the set of tensors dynamically promoted to the GPU.

\subsubsection{Dynamic Transfer Algorithm}

Algorithm~\ref{alg:system:dp-transfer} uses dynamic programming to minimize the latency of the current round. It traverses tensors in execution order while distinguishing whether the current partial schedule ends on the GPU or on the CPU. For tensors that are GPU-resident by default, the algorithm only needs to update the backend-switch cost. For tensors that are CPU-resident by default, besides the option of keeping them on the CPU, it also enumerates the start point of the most recent CPU$\rightarrow$GPU transfer and evaluates the corresponding exposed transfer cost. In this way, the dynamic program jointly captures execution cost, backend-switch overhead, and transfer/computation overlap within a unified optimization.

Empirically, the optimal transfer pattern varies by stage. During prefill, heavier computation and a larger overlap window make GPU execution more attractive, so more CPU-resident tensors tend to be promoted. During decode, the available overlap window is smaller, and the scheduler becomes correspondingly more selective, often yielding an intermittent promotion pattern. These results suggest that dynamic transfer should adapt to stage-specific execution characteristics rather than rely on a fixed policy.

In the worst case, the algorithm enumerates a legal transfer start point $j$ for each tensor that is CPU-resident by default, which yields $O(n^2)$ time complexity. The interval values $seg(j,i)$ can be precomputed in $O(n^2)$ time, so the overall runtime remains $O(n^2)$. Explicitly storing all interval values requires $O(n^2)$ space; if only dynamic-programming states and the information needed for backtracking are retained, auxiliary state can be reduced to $O(n)$. In practice, the overhead is usually lower because dynamic promotion is considered only for the subset of tensors that are CPU-resident under the static placement.

\subsubsection{Load-Aware Re-scheduling}

\system{} records the observed transfer and computation times of the current plan and compares them with the measurements used by the most recent scheduling decision. As summarized in Algorithm~\ref{alg:system:redecision}, rather than rerunning scheduling whenever load changes, \system{} triggers re-scheduling only when the measured deviation exceeds a threshold $\epsilon$ and sufficient time has elapsed since the previous re-scheduling event. This design is necessary because reconfiguration overhead is non-negligible relative to decode latency. For Qwen3-14B on an RTX 3060, TPOT is approximately 40~ms, whereas rebuilding the computation graph takes about 7~ms and rerunning the dynamic program takes about 9~ms. In our implementation, we therefore set $\epsilon$ to 15\% and enforce a minimum re-scheduling interval equal to five times the recent TPOT, thereby avoiding repeated re-optimization under short-term performance fluctuations. This threshold-based, rate-limited policy preserves load awareness while keeping reconfiguration overhead under control.

\begin{algorithm}[t]
\caption{Load-aware Re-scheduling}
\label{alg:system:redecision}
\begin{algorithmic}[1]
\STATE \textbf{Input:}
\STATE \hspace{1.5em}previous measurement snapshot $m^{prev}$;
\STATE \hspace{1.5em}current measurement snapshot $m^{cur}$;
\STATE \hspace{1.5em}deviation threshold $\epsilon$;
\STATE \hspace{1.5em}minimum re-scheduling interval $\tau$;
\STATE \hspace{1.5em}time of last re-scheduling $t^{last}$;
\STATE \hspace{1.5em}default placement $b$
\STATE \textbf{Output:}
\STATE \hspace{1.5em}runtime backend assignment $rb$

\STATE $d \gets \textsc{ComputeDeviation}(m^{prev}, m^{cur})$;

\IF{$d < \epsilon$ \textbf{or} $t^{now} - t^{last} < \tau$}
	\STATE \textbf{Return} previous assignment
\ELSE
	\STATE $(t^g, t^c, c, w) \gets m^{cur}$;
	\STATE $rb \gets \textsc{DynamicTransferScheduling}(b, t^g, t^c, c, w)$;
	\STATE $\textsc{BuildRuntime}(rb)$;
	\STATE Update $m^{prev} \gets m^{cur}$ and $t^{last} \gets t^{now}$;
	\STATE \textbf{Return} $rb$
\ENDIF
\end{algorithmic}
\end{algorithm}

%% file: chapters/5-evaluation.tex
\section{Evaluation}
\label{sec:evaluation}

\subsection{Experimental Setup}

\noindent\textbf{Hardware.}
We evaluate on two representative classes of consumer devices: a high-performance desktop platform and a mid-range laptop platform. Table~\ref{tab:hardware} summarizes the CPU, GPU, memory, and PCIe configurations of both systems. The laptop platform is constrained by thermal and power limits, making it difficult to sustain stable decode latency over long runs, whereas the desktop platform provides more stable sustained performance.

\begin{table}[t]
\centering
\caption{Hardware platforms in the evaluation.}
\label{tab:hardware}
\small
\setlength{\tabcolsep}{4pt}
\begin{tabular}{@{}lll@{}}
\toprule
  & Laptop & Desktop \\
\midrule
CPU & Intel i7-11800H & Intel i7-11700 \\
GPU & RTX 3060~(6GB) & RTX 4090~(24GB) \\
RAM & 32GB DDR4 & 64GB DDR4 \\
PCIe & Gen4 x16 & Gen4 x16 \\
\bottomrule
\end{tabular}
\end{table}

\noindent\textbf{Models.}
We evaluate \system{} across a range of state-of-the-art dense and MoE models with different parameter scales. On the RTX 3060 platform, we test GLM-Z1-9B~\cite{glmz1}, Qwen3-14B, Qwen3-30B-A3B~\cite{qwen3}, and GPT-OSS-20B~\cite{gpt-oss}. On the RTX 4090 platform, we test Llama3.1-70B~\cite{llama3}, Qwen3-Next-80B-A3B, Qwen3.5-122B-A10B~\cite{qwen3.5}, and GPT-OSS-120B. GLM-Z1-9B is evaluated in FP16, GPT-OSS models use the native MXFP4 format, and all remaining models use INT4 quantization.

\noindent\textbf{Baselines.}
We compare \system{} against three state-of-the-art baselines: llama.cpp~\cite{llama.cpp}, vLLM~\cite{vllm}, and KTransformers~\cite{ktransformers}. llama.cpp serves as the primary baseline and implementation foundation of \system{}, and represents a high-performance C++ runtime for practical heterogeneous inference. vLLM is widely used in server-oriented deployments and also provides support for single-GPU offloading. KTransformers targets MoE inference and already integrates SGLang~\cite{sglang} for GPU-side execution. Importantly, vLLM and KTransformers do not yet support the full model set considered in this paper. In particular, vLLM~(v0.18.0) does not currently support offloading for some recent models with customized attention implementations, and KTransformers~(v0.5.2) supports only a subset of MoE models because its custom CPU kernels do not support the MXFP4 format.

\noindent\textbf{Workloads and Metrics.}
We evaluate prefill throughput with prompt lengths ranging from 512 to 4096 tokens, and decode throughput with a prompt length of 2048 tokens and a generation length of 128 tokens. To study sustained-generation stability on the laptop platform, we further vary the output length from 32 to 2048 tokens and examine how decode throughput evolves over the course of generation. Beyond throughput, we also measure TPOT under simulated hardware pressure and compare GPU SM utilization and effective PCIe bandwidth to better understand the source of performance differences. All experiments use batch size 1, which matches the most common local-deployment setting on consumer devices. For fairness, all systems use the same chunk size for chunked prefill~(512 on the RTX 3060 platform and 2048 on the RTX 4090 platform), the same offloading policy that fills GPU memory as much as possible, and the same KV-cache size.

\subsection{End-to-End Performance}

\begin{figure*}[t]
\centering
\includegraphics[width=\linewidth]{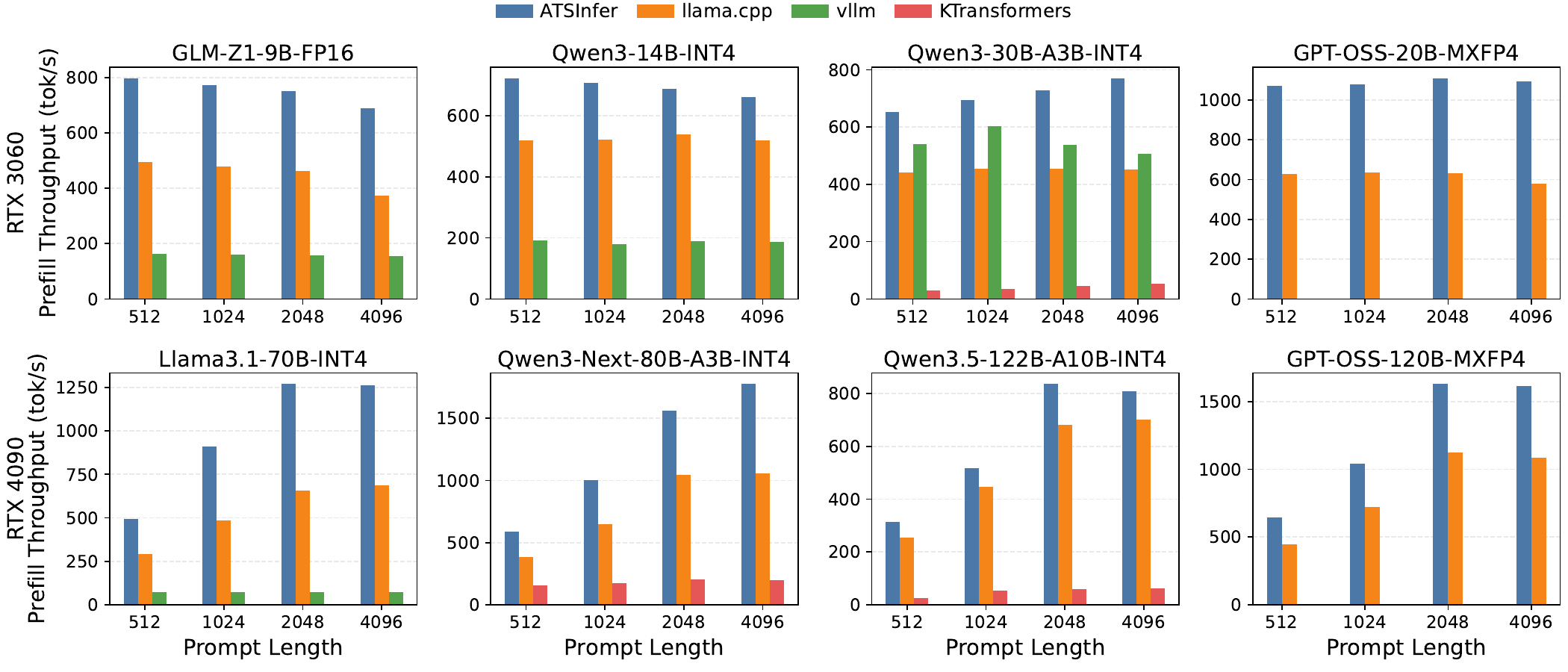}
\caption{Prefill throughput comparison between \system{} and other baselines on laptop and desktop platforms. Please note that vLLM~(v0.18.0) does not currently support offloading for Qwen3-Next, Qwen3.5, or GPT-OSS. KTransformers~(v0.5.2) supports the MoE models except GPT-OSS, because its custom CPU kernels do not yet support the MXFP4 format.}
\Description{A grid of bar charts comparing prefill throughput across models and platforms for \system{} and the baselines.}
\label{fig:eval:prefill-throughput}
\end{figure*}

\begin{figure}[t]
\centering
\includegraphics[width=\linewidth]{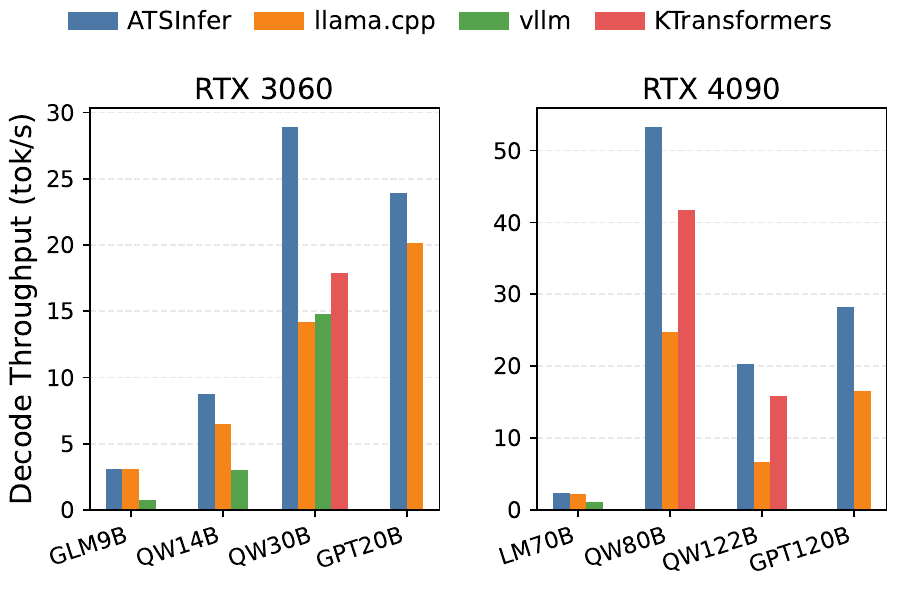}
\caption{Decode throughput comparison between \system{} and other baselines on laptop and desktop platforms.}
\Description{A grid of bar charts comparing decode throughput across models and platforms for \system and the baselines.}
\label{fig:eval:decode-throughput}
\end{figure}

\begin{figure}[t]
\centering
\includegraphics[width=\linewidth]{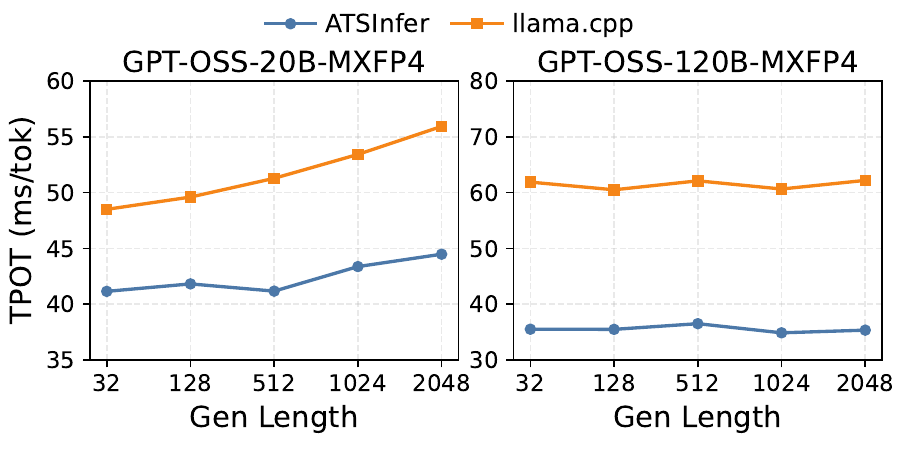}
\caption{Average decode TPOT as generation length increases on the laptop and desktop platforms.}
\Description{A plot showing average decode TPOT versus generation length for llama.cpp and \system{} on both the laptop and desktop platforms. The laptop curves increase over time, while the desktop curves remain comparatively stable.}
\label{fig:eval:decode-tpot-length}
\end{figure}

Figures~\ref{fig:eval:prefill-throughput} and~\ref{fig:eval:decode-throughput} compare the throughput of \system{} against the baselines in the prefill and decode phases, respectively.

In prefill, KTransformers delivers low throughput on consumer platforms because its optimized AMX-based CPU kernels are unavailable on consumer CPUs, resulting in substantial CPU-side overhead. vLLM also becomes less competitive as the activated parameter size increases, indicating that a fully GPU-centric strategy remains constrained by the PCIe bottleneck. Meanwhile, the ggml backend of llama.cpp relies on hard-coded thresholds during prefill to identify compute-intensive matrix-multiplication operators and place them on the GPU. By contrast, \system{} combines static tensor placement with dynamic transfer to schedule tensors automatically, and further uses CPU-GPU coordination to overlap transfer with computation more effectively than the serialized execution path in llama.cpp. As a result, \system{} improves prefill throughput over llama.cpp from 1.28$\times$ to 1.88$\times$ on the laptop platform and from 1.15$\times$ to 1.94$\times$ on the high-end desktop platform.

In decode, the computation is less intensive than in prefill, which allows some low-parallelism operators to be handled effectively by the CPU and makes KTransformers substantially stronger in decode than in prefill. On the models it supports, KTransformers consistently outperforms llama.cpp because expert-granularity offloading effectively selects a set of weights with relatively small average empirical performance density gaps between CPU and GPU, thereby incurring a smaller penalty from CPU execution. vLLM remains limited by the PCIe bottleneck inherent in GPU-centric offloading, and consequently performs well only for models with relatively small per-step weight transfer, such as Qwen3-30B-A3B~(about 1.0~GB), compared with Qwen3-14B~(about 4.9~GB) and GLM-Z1-9B~(about 13.5~GB). Guided by empirical performance density, \system{} makes placement decisions at tensor granularity and uses dynamic transfer to exploit otherwise idle PCIe bandwidth to relieve CPU pressure, thereby achieving the best decode performance overall. On the laptop platform, the maximum decode speedups of \system{} are 3.29$\times$ over llama.cpp, 4.35$\times$ over vLLM, and 3.15$\times$ over KTransformers. On the high-end desktop platform, the corresponding maximum speedups are 3.12$\times$, 2.03$\times$, and 1.33$\times$.

Figure~\ref{fig:eval:decode-tpot-length} shows how average decode TPOT changes with generation length on the laptop and desktop platforms. On the laptop platform, both llama.cpp and \system{} exhibit rising TPOT as generation continues, because thermal throttling prevents the device from sustaining its peak operating frequency over long runs. By contrast, the desktop platform provides more stable TPOT across different generation lengths.

\subsection{Load Adaptation}

\begin{figure}[t]
\centering
\includegraphics[width=\linewidth]{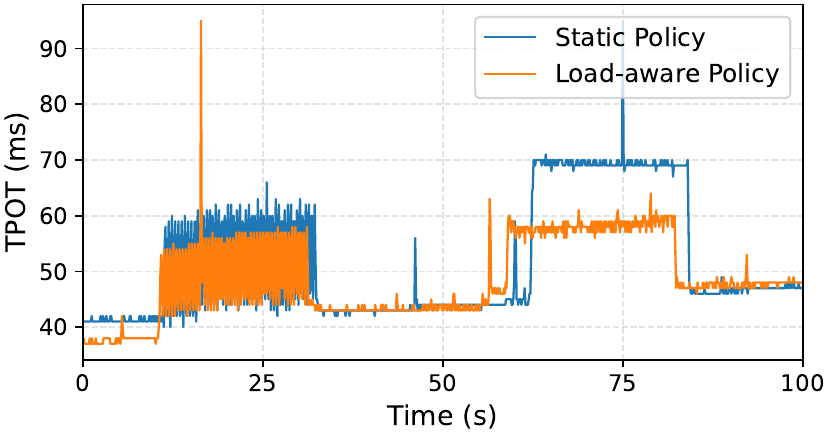}
\caption{Decode TPOT of static policy and load-aware policy under time-varying system load. CPU pressure is applied first, followed by GPU+PCIe pressure.}
\Description{A timeline plot showing decode TPOT under changing CPU, GPU, and PCIe load, comparing the adaptive behavior of \system{} with a baseline.}
\label{fig:eval:load-timeline}
\end{figure}

Motivated by the load variation discussed in Section~\ref{sec:motivation}, we next evaluate whether \system{} can adapt to realistic changes in CPU, GPU, and PCIe pressure, as illustrated in Figure~\ref{fig:eval:load-timeline}. We compare two policies: a static policy, which is manually tuned from offline profiling data, and a load-aware policy, which is generated online by the Dynamic Transfer Scheduling algorithm in Algorithm~\ref{alg:system:dp-transfer}. When CPU, GPU, and PCIe resources are perturbed by co-running applications, the static policy quickly becomes suboptimal, whereas the load-aware policy adjusts tensor transfer decisions according to current runtime conditions and thereby reduces performance degradation. As a result, compared with the static policy, the load-aware policy reduces the TPOT increase by 13\% under 60\% external CPU load and by 38\% under 60\% external GPU+PCIe load.

\subsection{Hardware Utilization}

\begin{figure}[t]
\centering
\includegraphics[width=\linewidth]{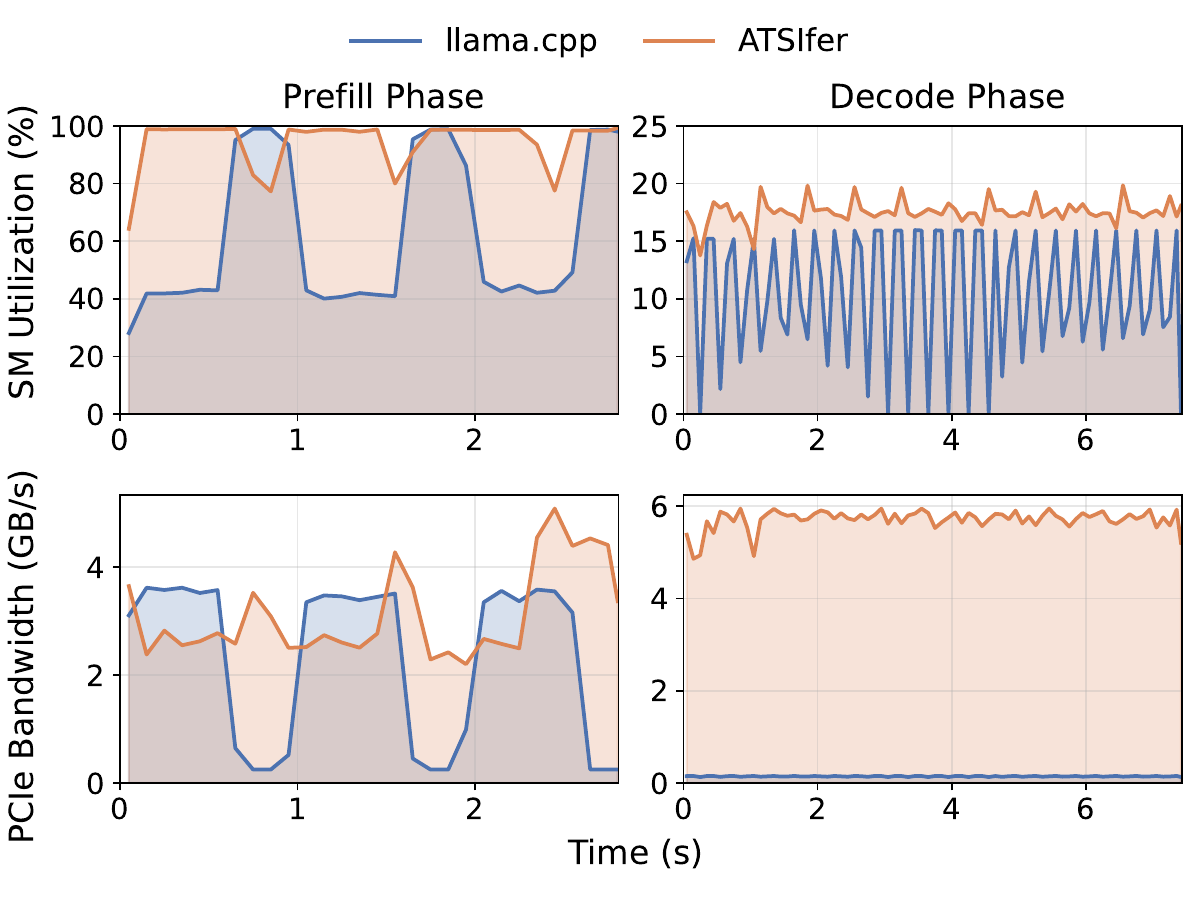}
\caption{Timeline of GPU SM utilization and effective PCIe bandwidth for \system{} and llama.cpp across the prefill and decode phases.}
\Description{A timeline plot comparing \system{} and llama.cpp across inference phases, showing GPU SM utilization and effective PCIe bandwidth over time. \system{} sustains higher GPU utilization during prefill and decode, and shows more persistent PCIe bandwidth usage during decode.}
\label{fig:eval:nsys-phases}
\end{figure}

Figure~\ref{fig:eval:nsys-phases} compares GPU SM utilization and effective PCIe bandwidth over time for \system{} and llama.cpp across different inference phases, using traces collected with NVIDIA Nsight Systems~\cite{nsight}.

During prefill, \system{} and llama.cpp keep a comparable amount of weights resident in GPU memory and transfer a similar total volume of data from CPU memory. The key difference is that \system{} benefits from its static tensor placement policy, which places more compute-intensive weights on the GPU, while also overlapping transfer and execution more effectively. As a result, \system{} sustains markedly higher GPU SM utilization during prefill, indicating that the GPU spends less time waiting on serialized transfer-compute interactions.

During decode, \system{} not only transfers intermediate activations, but also continuously moves selected CPU-resident weights into GPU memory according to runtime scheduling decisions. This design leads to more sustained PCIe bandwidth usage over time than in llama.cpp. In turn, the additional data movement increases PCIe bandwidth utilization and improves average GPU SM utilization by about 70\% relative to llama.cpp.

\subsection{Ablation Study}

\begin{figure}[t]
\centering
\includegraphics[width=\linewidth]{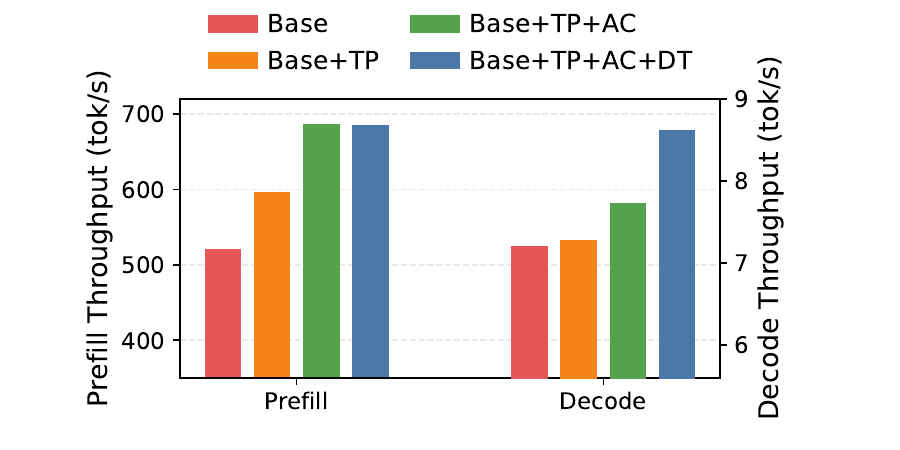}
\caption{Performance breakdown of \system{} on Qwen3-14B. The four configurations are \textit{base}~(llama.cpp), \textit{base+TP}, \textit{base+TP+AC}, and \textit{base+TP+AC+DT}~(\system{}), where TP denotes tensor placement, AC denotes async coordination, and DT denotes dynamic transfer.}
\Description{A bar chart showing the incremental performance impact of static tensor placement, asynchronous CPU-GPU coordination, and load-aware dynamic transfer on prefill and decode throughput for Qwen3-14B.}
\label{fig:eval:ablation-throughput}
\end{figure}

To better understand the contribution of each component in \system{}, we perform a detailed performance breakdown on Qwen3-14B with prompt length 1024 and generation length 128. Starting from llama.cpp, we incrementally add three mechanisms: Static Tensor Placement, asynchronous CPU-GPU Coordination, and Load-Aware Dynamic Transfer.

Figure~\ref{fig:eval:ablation-throughput} shows that the gains in prefill and decode arise from different sources. In prefill, the dominant improvements come from Static Tensor Placement and asynchronous CPU-GPU Coordination, with most of the benefit already realized before dynamic transfer is enabled. During decode, static placement alone provides only a modest gain, whereas the larger improvements come from asynchronous CPU-GPU Coordination and Load-Aware Dynamic Transfer. This trend is consistent with the utilization analysis above and indicates that runtime coordination and adaptation are more important in the latency-sensitive decode phase.

%% file: chapters/6-related-work.tex
\section{Related Work}
\label{sec:related-work}

\subsection{GPU-Centric and Hybrid LLM Inference}

Recent systems have made LLM inference increasingly practical under limited GPU memory, but they adopt different assumptions about the role of CPU in the execution path. GPU-centric systems such as FlexGen, DeepSpeed Offload, and Hugging Face Accelerate directly target memory-limited execution by moving weights or runtime state across a storage hierarchy while keeping GPU as the main compute backend~\cite{flexgen,huggingface,deepspeed}. Systems such as vLLM and SGLang are primarily designed for high-throughput server-side serving, but their memory-management and offloading mechanisms are also relevant to single-GPU deployment when model size exceeds device memory~\cite{vllm,sglang}. Across this line of work, GPU remains the dominant compute backend, and performance is fundamentally shaped by host-to-device bandwidth and by how effectively transfer latency can be hidden.

Hybrid systems move beyond this design by allowing CPU to participate directly in inference. \textsc{llama.cpp} shows that local hybrid execution can make LLM inference practical across a wide range of commodity devices~\cite{llama.cpp}. FlexInfer explores CPU-GPU cooperation to reduce transfer bottlenecks, while PowerInfer and LLM in a Flash exploit sparsity or hot-neuron behavior to retain part of the computation on the GPU while relying on CPU-side storage for the rest~\cite{flexinfer,powerinfer,llminaflash}. These systems demonstrate that hybrid execution is viable, but most still rely on layer-level partitioning, fixed placements, or coarse heuristics.

\system{} differs from this line of work in two key respects. First, it treats tensors rather than layers as the main scheduling unit, following the fine-grained opportunity highlighted in Section~\ref{sec:motivation}. Second, it combines offline static placement with runtime load-aware transfer, rather than relying on a single fixed partition or on GPU-centric transfer overlap alone.

\subsection{MoE-Aware Hybrid Inference}

Mixture-of-Experts models present an additional opportunity for hybrid inference because expert sparsity reduces the amount of model state and computation required for each token. Recent systems~\cite{ktransformers,hybrimoe,moelightning,dali,promoe,moe-infinity,adapmoe} exploit this structure to distribute storage and execution across CPU and GPU. Their results show that routing-aware placement and execution can make large MoE models practical on resource-constrained devices. Some of these systems also support online scheduling by adapting expert residency and swapping decisions to input-dependent changes in the activated-expert distribution~\cite{hybrimoe,moelightning,promoe,moe-infinity,adapmoe}. However, they do not explicitly account for time-varying hardware load, which materially affects performance on consumer devices. Their coarse scheduling granularity and lack of load-aware adaptation therefore limit performance in local deployment.

Importantly, \system{} applies the same optimizations to both dense and MoE models and does not yet incorporate MoE-specific optimization mechanisms. As a result, expert-residency and swapping strategies are complementary to \system{} and could potentially be combined with its tensor-granularity scheduling within experts to achieve further speedups.

\subsection{Compression and Model Reduction}

Model compression is a fundamental way to reduce the deployment cost of LLMs on consumer devices. Its primary objective is to reduce parameter size, memory traffic, and runtime storage demand while preserving model quality as much as possible.

Among existing approaches, low-bit quantization is the most widely used~\cite{smoothquant,awq,omniquant,gptq,llama.cpp}. By representing weights at lower precision, it substantially reduces both model size and transfer volume. In practice, 4-bit weight quantization is a common operating point for consumer-device deployment because it often offers a favorable balance between model accuracy and system efficiency~\cite{awq,gptq}. Beyond quantization, sparsity and pruning~\cite{dejavu,powerinfer,llminaflash,minference} reduce the number of parameters or activations involved in inference, thereby lowering computation, storage, and memory-access overheads. Knowledge distillation~\cite{distillingnn,kd_survey} reduces deployment cost from the model side by transferring the capability of a large model to a smaller one, which is typically easier to deploy under constrained memory and compute budgets.

%% file: chapters/7-conclusion.tex
\section{Conclusion}
\label{sec:conclusion}

In this paper, we presented \system{}, a system for efficient hybrid CPU-GPU inference of large language models on consumer devices. By combining tensor-granularity static placement, load-aware dynamic transfer, and asynchronous CPU-GPU coordination, \system{} achieves substantially higher throughput than existing baselines, with up to 1.94$\times$ higher prefill throughput and 3.29$\times$ higher decode throughput than llama.cpp. These optimizations mitigate key bottlenecks in coarse-grained offloading and load-unaware scheduling under tight memory and bandwidth constraints. Overall, \system{} helps make local LLM deployment on resource-limited consumer devices more practical and responsive.